\documentclass[12pt]{article}

\usepackage{amsmath}
\usepackage{amssymb}

\usepackage{graphicx}

\usepackage{cite}

\usepackage{color} 

\usepackage[labelfont=bf,labelsep=period,justification=raggedright]{caption}

\makeatletter
\renewcommand{\@biblabel}[1]{\quad#1.}
\makeatother

\date{}

\begin{document}

\begin{flushleft}
{\Large
\textbf{Fast Moving Sampling Designs in Temporal Networks}
}
\\
Steven K. Thompson$^{1}$, 
\\
\bf{1} 
 Department of Statistics and Actuarial Science, Simon Fraser
 University, Burnaby, BC, Canada
\\
$\ast$ E-mail: thompson@sfu.ca
\end{flushleft}

\section*{Abstract}


\section*{Introduction}

In a study related to this one \cite{thompson2015interventions} I set
up a temporal network simulation environment for evaluating network
intervention strategies.  A network intervention strategy consists of
a sampling design to select nodes in the network.  An intervention is
applied to nodes in the sample for the purpose of changing the wider
network in some desired way.  The network intervention strategies can
represent natural agents such as viruses that spread in the network,
programs to prevent or reduce the virus spread, and the agency of
individual nodes, such as people, in forming and dissolving the links
that create, maintain or change the network.  The present paper
examines idealized versions of the sampling designs used to that
study.  The purpose is to better understand the natural and human
network designs in real situations and to provide a simple inference
of design-based properties that in turn measure properties of the
time-changing network.  The designs use link tracing and sometimes
other probabilistic procedures to add units to the sample and have an
ongoing attrition process by which units are removed from the sample.

Properties of with-replacement random walks in graphs and their Markov
chain Monte Carlo modifications are described
\cite{boyd2004fastest}. Eigenvector values of random walk designs have
been used as measures of centrality in social networks since
\cite{bonacich1972factoring}, \cite{bonacich2007some}.  Limiting
selection probabilities of random walk in a network were proposed as a
ranking measure for web pages returned by a search engine were
proposed in \cite{brin2012reprint}. The difficulties of obtaining a
random sample of web pages is discussed in
\cite{henzinger1999measuring}.  Use of random walk design to reach all
nodes in a network as fast as possible is taken up in
\cite{avin2008explore}.  A ``forest fire'' sampling design in
\cite{leskovec2005graphs}, based on a network creation procedure in
\cite{leskovec2006sampling}, traces a random number of links out and
adds nodes without replacement. Links in are back-traced with a lower
probability.  The purpose is to obtain a sample that, somewhat like a
large random sample, captures characteristics of the complete network.

The designs of this paper offer an increased flexibility, have both an
acquisition and a attrition process, giving properties of stochastic
stationarity where desired.  In temporarily changing networks the
moving average and exponential smoothing measures allow the
with-replacement versions of the designs to
continuously approximate their own inclusion properties, which in turn
can be used to highlight temporally high-interest areas of the
network.  Thes measures are much more concentrated than are the
eigenvector values of random walks.  Without-replacement versions of
these designs, or modifications allowing with-replacement with lower
probability or  only when stuck can be used to reach new areas of the
changing network over time.  

\section*{Methods}

\subsection*{Fast moving designs for design-based inference}

Since the purpose of the simulations is to assess the effectiveness of
intervention designs and since the virus in the simulations has a
tendency to explode unexpectedly into new areas, it is some interest
to try to anticipate where it might potentially explode into next.
One possibility is to keep a running tabulation on connected components and note
when the virus spread has intersected with a component.  However,
components are continually changing and a component might be
technically connected but thinly so from one part to another.  That is
there may be few paths from one part of a component to another.
Another possibility is to look at the population graph as if frozen in
time and calculate the expected increase of virus prevalence at the
next step using the number and configuration of links our from the
current virus sample.  At the same time the expected decrease is
calculated using the stage specific node mortality rates for the
virus-infected nodes.  Birth, death, immigration, and emigration rates
can be weighed in also.  The difference between the rates of increase
and decrease gives a partial measure of the expected net change.  Such
a calculation is conditional on the network configuration at the given
time $t$.  What is missing from the calculation is the social movement
of individuals, the drift of social groups with respect to each other,
and the potentials for new formations of links after those movements,
as well as deletions of current links.  

A different type of measure uses idealized designs that are similar to
the designs by which the virus spreads but faster moving and faster
mixing.  There are many possible variations on such designs.  Among
the simplest traces links out with independent Bernoulli probabilities
$p$ per link.  The value of $p$ here tends to be much higher than the
transmission rates of HIV and higher than the realistic tracing rates
of seek and treat intervention designs.  The simplest of these designs
uses equal tracing rates.  We consider several of these idealized
types.

\textbf{Design 1}.  At time $t$ links out from the current sample are traced
with equal probability $p_t$, independently.  The target sample size
is $\nu$.  Sample size just before additions and removals at time $t$ is $n_{t-1}$.  The rate at which
nodes in the sample are removed from the sample is denoted $r_t$, that
is, for any node in the sample, $r_t$ is the probability
it is removed at that time step.  To maintain a sample size
stochastically around $\nu$ either $p_t$ or $r_t$ or both are adjusted
based on current
sample size, that is, on $n_{t-1}$.  With ``back control'', let $p_t$ be constant and $r_t = (n_t - \nu)/n_t$ if
$n_t - \nu > 0$ and $r_t = 0$ otherwise.  

With ``front control'', let removal rate 
$r_t$ be constant and adjust tracing rate $p_t$ based on $n_{t-1}$ to
add stochastically any shortfall from target sample size $\nu$.  Since
some nodes not in the sample may be linked from more than one node in
the sample, the expected number of nodes added for a fixed rate $p_t$ is
\[
\sum_{ \{i: i\notin s_{t-1}\}}\left[1 - \prod_{\{j:j\in s_{t-1}, e(j,i) \in E_{t-1}\}}(1-p_t)  
\right]
\]
As an approximation to stochastically maintain the sample size around
the target size, set the tracing rate to be 
$p_t = (\nu - n_{t-1})/\#E_{s+}$ if $\nu - n_{t-1}  > 0$
and $p_t = 0$ otherwise, where $\#E_{s+}$ is the number of links out
from the sample.  That is, $\#E_{s+} = \#\{e(i,j)\in E_{t-1}: i \in s_{t-1}, j
\notin s_{t-1}\}$.  

It is sometimes of interest to have fixed rates $p$ and $r$ of
link-tracing and node removal.  In such a case the stochastic sample
size $n_t$ stochastically reflects the density of nodes in the
network.  For some types of dynamic networks, such as those with
preferential attachment depensation in link formation, the
fluctuations in sample size under fixed rates can be very pronounced.  

In addition, at time $t$ nodes not in the sample may be added
independently at random to the sample with some small probability
$d_t$, so that the expected number of nodes added this way is $d_t(N_t
- n_t)$.  

This design is without-replacement in that a node currently in the
sample just before time $t$ is not re-selected at time $t$.  It is
with-replacement, however, in the sense that a node previously in the
sample that has been removed can be re-selected at any time.  

A design of this type can be modified so that probabilities of
selection and removal depend on node or link values.  

\textbf{Design 2}.  This design is much like the first but is
with-replacement even of units currently in the sample.  At time $t$ a
unit $i$ can be selected even though it is already in $s_{t-1}$, which
is the sample as sampling first starts at time $t$.  Any unit in
$s_{t-1}$ can be re-selected only once at time $t$.  This design is
similar in its properties to Design 1, but concentrates farther into
the most link dense, connected parts of the population.

\textbf{Design 3}.  This design concentrates still farther by letting
the sampling at time $t$ be completely with-replacement.  With each
unit $i$ we tabulate $m_t(i)$, the number of times it has been
selected at time $t$.  

\textbf{Design 4}.  Here we let Design 1 be completely
without-replacement, so that any unit that is currently or has ever
been previously in the sample is not eligible for re-selection.  In
contrast to the first three designs which tend to concentrate their
effort in highly connected parts of the population, the
without-replacement design, even though it is tracing links, tends to
spread farther and farther out, reminiscent of a random sample, as
time goes on.  The design can get stuck, with no links out to units
that have not been already sampled at some time.  A small random
selection probability will let it eventually jump to new nodes.  A
variation of this design allows with-replacement sampling only when
link tracing is stuck.  It is also possible to have a steady, small
probability of with-replacement sampling.  

\textbf{Design 5}.  This design is specifically for use with an
epidemic such as HIV.  In this design link-tracing continually begins
with nodes infected with the virus.  At time $t$ independent Bernoulli
sampling selects infected nodes with probability $p_0$.  Link tracing
of nodes already in the sample (from time $t-1$) independently with
probability $p_l$.  There is a constant attrition rate $r$ removing
nodes from the sample.  In this way selection and attrition come into
dynamic, fluctuating balance.  No set limit is placed on the number of
steps links are followed out from infected nodes, but attrition
stochastically limits that number to a few.  The sample thus
highlights uninfected nodes at most immediate risk for infection. An
uninfected node with more network links or paths from the current
virus sample tends to be in the sample more often than a node that
is connected to the virus sample by fewer links or paths.  Variations
on this design have set rules on number of steps out, for example
fixed limit one step or two steps of tracing from an infected node.

\textbf{Design 6}.  In each of the above designs the sampling is
limited to a single round, that is, a single wave of selections at
time $t$.  Design 6 takes any of those designs and lets it run for
some number $k$ of waves at time $t$.  Thus, for designs having stable
limiting distribution properties, including limiting inclusion
probabilities for each node, that limit can be approached at each time
step.  As the network itself changes, the limits change at each time
step.  

All of these designs can alternatively be considered a single type of
network design with different parameter values corresponding to
acquisition and removal rates, existence of sample size targets,
with-replacement and without-replacement types and rates, recruiting activeness
depending on factors such as time in sample, coupons held, numbers
of waves at a time step, and similar characteristics.  Natural designs
such as used by a virus to spread in a population, and practical
designs such as seek and treat designs for spreading tests and
interventions and respondent-driven designs for assessing risk
behaviors in a population, are modeled as designs of these types
having special characteristics.  One design may interact with another
through competition or mutual enhancement, one design may modify its
parameter values to adapt to another, and the effect of intervention
designs can include changing the network structure of the population. 

\textbf{Design 0}.  Random walk designs in the dynamic network are
collectively called Design 0.  At time $t-1$ the random walk sits on
some node, say node $i$, so that the indicator variable $Z_{t-1}(i)=1$
and for all other nodes $j \ne i$, $Z_{t-1}(j)=0$.  Let $d_t(i)$ be
the number of nodes to which node $i$ connects.  That is, $d_t(i)$ is
the out-degree of node $i$.  The random walk selects one of these
links at random, with probability $1/d_t(i)$, and traces that to its
destination node, say node $j$.  The random walk now sits on node $j$,
so that $Z_t(j) = 1$ and $Z_t(i)=0$.  Sampling is with-replacement, so
that the walk can move back to node $i$ at any time that is is
connected to the current walk node.  If the random walk has a positive
probability of staying at the node where it is at a time step, it is
called a ``lazy'' random walk.  The random walk design can be modified
to allow weights on links with probability of following a link
depending on its weight relative to the other links out.
 
When a random walk design is run in a network that changes with time,
some behaviors are seen that do not occur with a random walk in a
static network.  In a static network, a random walk is stuck in the
single connected component in which it starts, unless it is allowed to
make a jump at from time to time to a randomly selected unit.  In the
dynamic graph setting, a random walk might be stuck for some time in
an isolated component.  Bot over time that component might connect
with another and a new component formed, so the random walk moves in
the new component.  The random walk might move to a node whose links
are subsequently deleted, so that it is stuck on that single node
until a new link forms, unless it is allowed to make a random jump at
that point.  Further, the node on which the walk sits might be deleted
from the graph at time $t$, in which case the walk ends.  The design
might be modified to start a new walk at each such event.

When $\nu$ random walks are run independently in the same graph, we
can view it as a single design of our general type as follows.  At any
time $t-1$ any node $i$ in the population has associated with it a
value $M_{t-1}(i)$ representing the number of walks that sit on that
node at that time;  $0 \le M_{t-1}(i) \le \nu$ and $\sum_{i \in U_t}
M_{t-1}(i) = \nu$.  At time $t$ for node $i$, $M_{t-1}(i)$ independent multinomial
selections are made, the result of each selection being a transfer of
one count transfered from node $i$ to one of its linked nodes or else
remaining at node $i$.  When a count is transferred to a linked node
$j$, $M_{t}(j)$ is incremented by 1 and $M_{t}(i)$ is decremented by
1.  Typically, random jumps are allowed, and are automatic when a node
having one or more walks on it is deleted.  A random jump means that a
node $j$ is selected at random from the population and 1 is added to $M_t(j)$.

\subsection*{Flame rank}

Any of the fast mixing designs of this section can be averaged or
smoothed to estimate design or network properties.  Let $Z_t(i) = 1$
if node $i$ is in the sample at time $t$ and $Z_t(i)=0$ otherwise, so
that the random collection $\{Z_t(i), i \in U_t\}$ are the sample
inclusion indicators for the nodes in the current population.   

With a static graph, for which $G = \{U, E\}$ does not change with time
but the sampling process $\{S_t\}$ progresses indefinitely in time with a stationary
distribution, the cumulative means $(1/T) \sum_{t=0}^T Z_t(i)$
converge in probability to fixed sample inclusion probabilities
$\pi_i$ as $T \rightarrow \infty$, for $i = 1, ..., N$.  

The limiting inclusion probabilities and associated stationary
distribution are particularly well understood for the random walk
design in a static graph, where the $(\pi_1, ..., \pi_n)$ is the first
eigenvector for the adjacency matrix of the graph, associated with
eigenvector $\lambda =1$, and the second eigenvector is a measure of
how fast the design mixes to that stationary distribution.  

The dynamic graph case has had much less study and is less well
understood, even for the random walk.

In the dynamic case each node has a sample inclusion indicator
variable $Z_i$ providing a time series of values 0 and 1 for each time
step while the node exists.  

Let $t_{oi}$ be the time that node $i$ came into existence or entered
the population.  At time $t$ the cumulative mean
$(1/t-t_0)\sum_{s=t_0}^t Z_s(i)$ represents the proportion of time the
node has been in the sample.  Because of changes in the network
structure during that time, the cumulative mean might not be a good
approximation for the current probability that node $i$ is included in
the sample.  

For the dynamic setting we calculate for each node a moving average
value, which averages the indicator variables for the most recent
values, back to some lag $L$.  That is, $v_t(i) = \sum_{j = 0}^L
\theta_j Z_{t-j}(i)$, with $\sum_{j=0}^L \theta_j =1$;  for instance
with equal weights $\theta_j = 1/L$ or half-normal weights $\theta_j =
e^{-(j^2)/2\sigma^2}/\sum_{k=0}^Le^{-(k^2)/2\sigma^2}$.  For
simplicity and computational efficiency we are using most often an
exponentially weighted moving average 
\[
v_t(i) = \lambda v_{t-1}(i) + (1-\lambda) Z_t(i) 
\] 
which corresponds to decreasing weights $\theta_j = \lambda^j$ for lag
$j$ going indefinitely into the past, with $|\lambda| < 1$.  

On one level we are using the fast moving design for a simple
design-based inference telling us only about the design itself.  That
is, the design is estimating its own unit inclusion probabilities in a
current interval of time.  But in doing so, current dynamic properties of the
network are also being estimated.  By its nature the design produces a
sample process that spends a
lot of its time with units that are well connected.  A unit spending a
lot of recent time in the sample is likely to be reachable by many
sample paths from other units, including direct links and paths of
length more than one.  Conversely, that unit has a lot of paths out to
other units.  A unit with a high current value of the moving average
$v_t(i)$ is a unit that, if it is sparked by an infection of a virus,
is likely to result in a flame spreading to other units with the
initial high transmission rate of early infection stage.  Visually, in
the dynamic network simulations one sees areas of high $v_t$ value,
colored in shades of red, that are areas of above average probability
of seeing a virus spread explosion.  For this reason we refer to
measures such as $v_t$ with the free-style link tracing set designs as
``flame rank.''

\section*{Results and Discussion}

This study uses fast-moving idealized temporal network sampling
designs to understand and anticipate the natural of natural network
designs, such as those of infectious agents, and intervention designs
used by human organizations and individuals to counteract those.  The
virus HIV adapts to an uneven, time-varying human sexual network by
having a high transmission rate in the short early stage of infection and a low
rate in the long chronic stage.  The high early rate enables explosive
spread in dense network clusters and, because of the trade-off between
transmission rate and mortality rate of host, allows long survival
between temporal clusters allowing some strains of virus to persist
until another cluster is reached.  

The flexible designs such as designs 1-3 in this paper spend much of
their time in the areas of the network in with such explosions can
occur.  By averaging the inclusion indicators of these designs using
moving average or exponential smoothing techniques, the designs
provide a measure that highlights areas in which virus would spread
fast if ignited by one or more initial infections.  This useful in
visual interpretation of simulation studies on intervention
strategies, and suggests that designs of similar types could prove
useful in distribution real-world interventions through temporally
changing networks.


\section*{Acknowledgments}
This work is supported by the Natural Science and
  Engineering Research Council of Canada.  

\bibliography{refs2015r}



\end{document}